\documentclass[aps,pra,twocolumn,showpacs,floatfix]{revtex4}
\usepackage{bm}
\usepackage{graphicx} 
\usepackage{amsmath}
\usepackage{epstopdf}
\usepackage{dsfont}

\begin{document}
\title{Can Nitric Oxide be Evaporatively Cooled in its Ground State?}

\author{Lucie D. Augustovi\v{c}ov\'{a}}
\altaffiliation{Alternative address: Charles University, Faculty of Mathematics and Physics, Department of Chemical Physics and Optics, Ke Karlovu 3, CZ-12116 Prague 2, Czech Republic}

\author{John L. Bohn}
\affiliation{JILA, NIST, and Department of Physics, University of Colorado, Boulder, CO 80309-0440, USA} 

\date{\today}

\begin{abstract}
Cold collisions of $^{14}$N$^{16}$O molecules in the $^{2}\Pi_{1/2}$ ground state, subject to electric and magnetic fields, are
investigated.  It is found that elastic collision rates significantly exceed state-changing inelastic rates only at temperatures above 0.5 K at laboratory strength fields.  It is found, however, that in very large fields $> 10^{4}$ V/cm, inelastic rates can be somewhat suppressed.  Magnetic fields have negligible influence on scattering for this nearly non-magnetic state.

\end{abstract}


\maketitle


\section{Introduction}
Progress in the production of ultracold molecules continues to accelerate, thanks especially to very recent experimental developments such as molecular laser cooling \cite{Shuman10_Nat,Barry14_Nat,Yeo15_PRL,Kozyryev16_CPC,Chae17_NJP,Anderegg17_preprint}  and buffer-gas-loaded beams \cite{Hutzler12_ChemRev}.  The former technique entails the real possibility of producing sub-$\mu$K molecular samples, although it is at present limited to a particular subset of molecules.  By contrast, the buffer-gas-loaded beams make for very general sources, but are still limited to sample temperatures of hundreds of mK.  At such temperatures the molecules are merely ``very cold,'' not yet ultracold.  The true ultracold regime would produce molecules with such low translational temperature that they collide in only a single partial wave, thus realizing the ``ultimate molecular beam experiment'' -- that is, state preparation in a single quantum number of all degrees of freedom, including the partial wave of relative motion of reactants \cite{Ospelkaus10_Sci}.

Thus for a number of chemically relevant species, notably radicals such as OH, CH, NH, and NO, additional cooling seems desirable, and may potentially be achieved by evaporative cooling.  This well-established technique selectively removes the highest-energy atoms from the sample and allows the rest to re-thermalize at a lower temperature.  It requires a relatively high elastic collision rate and at the same time a relatively low rate of inelastic or chemical collisions.  The technique works well for alkali atoms, where it was noted long ago that the ratio of elastic to inelastic collision rates should exceed $\sim 100$ for evaporation to be effective \cite{Monroe93_PRL}.  

For molecules, it has been somewhat problematic to achieve this ratio, owing to the rotational structure and dipole moments of the molecules, which provide many avenues for re-arranging energy and angular momentum, thus propelling molecules from trappable, weak-field-seeking states, to untrapped strong-field-seeking states.  Evaporative cooling is not completely out of the question for molecular radicals, however; evidence for evaporative cooling has been reported in OH \cite{Stuhl12,Reens17_preprint}.

Inelastic rates for many molecules have been studied at low temperatures in this context, as recently reviewed  in \cite{Bala16_JCP}.  Thus far, to our knowledge no one has investigated the situation for the important atmospheric radical NO, which could be tamed by  Stark deceleration \cite{Vogels15_Sci} or other means \cite{Bichsel07_PRA,Elioff03_Sci}.  Accordingly, here we undertake estimates of elastic and inelastic collisions of NO radicals at low temperature, in their $^2\Pi_{1/2}$ ground state.  We focus on weak-electric-field seeking states in the upper component of the zero-field parity doublet, as these would prove electrostatically trappable.  Inelastic collisions are then those that leave one or both molecules in the untrapped- strong-field-seeking state after the collision.  

In general, we find that inelastic collision rates are large for NO molecules in their $^2\Pi_{1/2}$ state, likely preventing evaporative cooling to temperatures below tens of mK for this state.  The rates are seen to vary slightly upon application of an electric field, but remain large for typically achievable laboratory fields $< 10^4$ V/cm.  We note, however, a lowering of the inelastic rates for fields higher than this.  We note also that magnetic fields have little effect on the $^2\Pi_{1/2}$ state, as it possesses a minuscule magnetic moment.


\section{Theory}

\subsection{Molecular Hamiltonian}
A single  $^2\Pi$ molecule, treated as a rigid rotor and immersed in electric and magnetic fields can be described by the following effective Hamiltonian
\begin{eqnarray}
\label{effHamiltonian}
H = H_{\rm SO} + H_{\rm ROT} + H_{\rm SR} + H_{\rm HFS} + H_{\rm S} + H_{\rm Z},
\end{eqnarray}
where the individual terms represent the electronic spin-orbit coupling
$H_{\rm SO}$,
rotational Hamiltonian
$H_{\rm ROT}$, 
spin-rotation coupling
$H_{\rm SR}$,
hyperfine structure
$H_{\rm HFS}$, 
and Stark and Zeeman Hamiltonians
$H_{\rm S}$, $H_{\rm Z}$.
For $^{14}$NO in its lowest vibronic state the spin-orbit coupling constant $A_{\rm SO}=123.146$ cm$^{-1}$ and the rotational constant $B_{\rm ROT}=1.696$ cm$^{-1}$ \cite{Varberg}, thus satisfying the requirement $A_{\rm SO}\Lambda \gg B_{\rm ROT}J$ for the Hund's case (a) to be a good representation.
In the absence of $\Lambda$-doubling, each $J,M$ angular momentum  state is doubly degenerate in the projection $\Omega$ of the angular momentum on the molecular axis,
\begin{align}
\begin{split}
\label{nonparbasis}
|^2\Pi^{\pm}_{3/2}\rangle&=|n,\Lambda=\pm1,S,\Sigma=\pm1/2\rangle|\Omega=\pm3/2,J,M\rangle \\
|^2\Pi^{\pm}_{1/2}\rangle&=|n,\Lambda=\pm1,S,\Sigma=\mp1/2\rangle|\Omega=\pm1/2,J,M\rangle\,,
\end{split}
\end{align}
where $n$ refers to any other unspecified quantum numbers.
In the presence of $H_{\rm SO}$ and $H_{\rm SR}$, the appropriate zero-field energy eigenstates are also eigenstates of parity $p= \pm$,
\begin{align}
|^2\Pi_{3/2}p\rangle&= \frac{1}{\sqrt{2}}\left(|^2\Pi^{+}_{3/2}\rangle +p(-1)^{J-S} |^2\Pi^-_{3/2}\rangle\right) \\
|^2\Pi_{1/2}p\rangle&= \frac{1}{\sqrt{2}}\left(|^2\Pi^{+}_{1/2}\rangle +p(-1)^{J-S} |^2\Pi^-_{1/2}\rangle\right)\,.
\label{paritydef}
\end{align}
Of these states, the $\Omega=1/2$ is the ground state of NO, with the lowest-lying $\Omega=3/2$ around 170\,K higher in energy.  The dominant inelastic processes at ultralow temperatures will be state-changing collisions within the $J=1/2$, $\Omega=1/2$ manifold,  which we will focus on here.

Moreover, the $^{14}$N nucleus has spin $I=1$ (spin of $^{16}$O nucleus is zero), whereby the appropriate quantum numbers are those of the hyperfine interaction, where $\vec{J}$ and $\vec{I}$ are coupled to form $\vec{F}$ in the lab frame. Thus the $J=1/2$ level of interest here splits into two hyperfine levels, $F=1/2$ and $F=3/2$. So we work in the basis set $|\eta, \Omega, J, I, F, M_F;p\rangle$, where $M_F$ is $F$'s projection onto the laboratory axis and $\eta$ is a general index which represents all other considered quantum numbers.  For simplicity in the following, we will abbreviate this state vector as $|FM_F,p\rangle$.  
We adopt the hyperfine levels for the $^2\Pi_{1/2}, J = 1/2$ rotational level from Ref. \cite{Meerts}.

In an applied homogeneous electric field, $\vec{\mathcal{E}}$, whose direction specifies the space-fixed $z$-axis, the effective Hamiltonian is augmented by the Stark-effect term $H_{\rm S} = -\vec{d_s}\cdot\vec{\mathcal{E}}$, whose matrix elements are given elsewhere \cite{Avdeenkov02,Ticknor05}.  The electric dipole moment of NO in its ground state is relatively small, $d = 0.15872$ Debye \cite{CRC}.  For this reason, fairly large electric fields are required to influence collisions of NO molecules, as we will see below.

Finally, in the presence of an external magnetic field the molecule experiences a Zeeman interaction with Hamiltonian $H_{\rm Z} = -\vec{\mu}\cdot\vec{B}$.  However, in the $\Omega=1/2$ state, contributions to the molecular magnetic moment due to the electron's orbital motion and spin nearly cancel, leaving this state with a negligible $g$-factor $\sim 0.0007$ \cite{Mizushima55_PR}.  We therefore neglect this interaction in most of what follows.

\begin{figure}[h]
\includegraphics[width=0.47\textwidth]{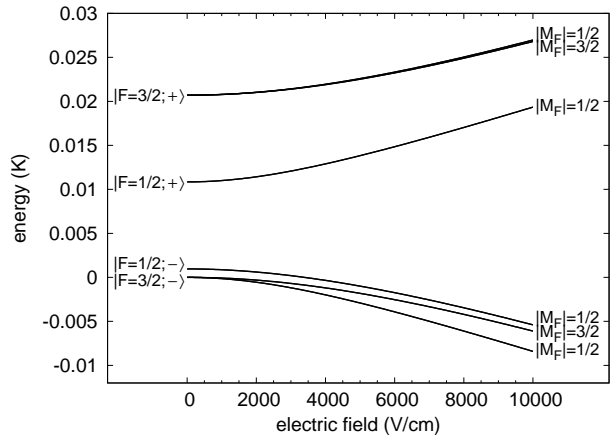}
\caption{Stark energies of the hyperfine $|{\tilde F},M_F,+/-\rangle$ and $\Lambda$-doublet levels for $J=1/2$ of the $^2\Pi_{1/2}$ state of the NO molecule.}
\label{Starkenergies}
\end{figure}

Fig. \ref{Starkenergies} demonstrates the Stark effect for a single NO molecule in the $^2\Pi_{1/2}$ ground state.
States of opposite parity repel as the field is increased, and the Stark effect transitions from quadratic to linear at a characteristic electric field $\mathcal{E}_0=\Delta_{\Lambda}/2d \sim 2500$ V/cm. The states in this figure are labeled by the quantum numbers ${\tilde F}$ and $|M_F|$, noting that $M_F$ referred to the electric field axis is a good quantum number, but $F$ is not; and that $\pm M_F$ states remain degenerate in an electric field.  Those states whose energies rise with electric field are the weak-field-seeking states that can be trapped electrostatically.  Their collisions that result in molecules in the lower-energy, high-field seeking states are the collisions that the experiment wishes to avoid, at least for evaporative cooling purposes.

\subsection{Molecular Scattering}

In general, collisions that remove the molecules from the weak-field-seeking states can take two forms: inelastic collisions from the upper to lower states depicted in Figure \ref{Starkenergies}; or else chemical reactions.  Taken together, these processes -- referred to collectively as ``quenching'' -- are generally indistinguishable, resulting only in loss of molecules from the trap, unless products are detected.  From the thermodynamic point of view, NO molecules are certainly exoergic in collision, and can undergo the reaction  2\,NO $\rightarrow$ N$_2$ + O$_2$ with  enthalpy change ${\scriptstyle \Delta} H = -180$ kJ/mol, thus releasing $21,000$ K of kinetic energy, a devastating blow to trapping.  However, from the kinetics point of view, this reaction is rendered unlikely due to the very high activation energy 209 kJ/mol $\doteq 25,000$ K \cite{Zhou}. Reaction probability will therefore be taken as negligible, although of course it could conceivably be enhanced by resonant tunneling \cite{Bodo04_JPB}.  

We therefore replace the short-range physics by a hard wall boundary condition at $R=30\,a_0$.  State-changing physics is then described by the long-range van der Waals and dipole-dipole interactions.   The former is taken to be isotropic and is given by 
$-C_6/R^6$, where we take $C_6=35.2 \,E_h\,a_0^6$ as a lower estimate of actual $C_6$ obtained from the London formula, using the mean static static polarizability for NO \cite{Bridge}.
The long-range dipole-dipole interaction is given by
\begin{align}
\begin{split}
V_{\rm dd}(\vec{R}) = &- \frac{3\,(\hat{R}\cdot\vec{d_1})(\hat{R}\cdot\vec{d_2})-\vec{d_1}\cdot\vec{d_2}}{4\pi\varepsilon_0\,R^3} \\
		    = &-\frac{\sqrt{6}}{4\pi\varepsilon_0\,R^3}\sum_{q=-2}^2 (-1)^{q} C_{q}^2({\hat R})  (d_1\otimes d_2)_{-q}^2
\label{Vdd}
\end{split}
\end{align}
where $\vec{R}=R\hat{R}$ is the intermolecular separation vector in relative coordinates, and $(d_1\otimes d_2)_{-q}^2$ is the $-q$-component of the compound irreducible second rank tensor product of the 1-rank tensors that act as the electric dipole moment operators on the individual variables of molecules 1 and 2. 

Basis functions for scattering calculations consist of states of the separated molecules, plus partial waves in the expansion of the relative motion.  These states will be denoted  $|{\tilde F}_1,M_{F1};p_1\rangle\!\rangle|{\tilde F}_2,M_{F2};p_2\rangle\!\rangle|L,M_L\rangle$, where the ${\tilde F}$ notation is used to emphasize that these states are eigenstates of the molecules in an electric field, where $F$ is not, strictly, a good quantum number.
For simplicity in the following, we will employ the shorthand notation
$|{\tilde F},M_F;p\rangle\!\rangle \equiv |\eta, \Omega, J, I, {\tilde F}, M_F;p\rangle$ for individual molecules. Basis functions symmetrized under particle exchange are given by (for ${\tilde F}_1\neq {\tilde F}_2$ or $M_{F1}\neq M_{F1}$ or $p_1\neq p_2$) 
\begin{align}
\begin{split}
|{\tilde F}_1&,M_{F1};p_1\rangle\!\rangle|{\tilde F}_2,M_{F2};p_2\rangle\!\rangle|L,M_L\rangle_S = \\
&\frac{1}{\sqrt{2}} \left\{ |{\tilde F}_1,M_{F1};p_1\rangle\!\rangle|{\tilde F}_2,M_{F2};p_2\rangle\!\rangle|L,M_L\rangle \right. \\
& \pm \left. (-1)^{L} |{\tilde F}_2,M_{F2};p_2\rangle\!\rangle|{\tilde F}_1,M_{F1};p_1\rangle\!\rangle|L,M_L\rangle \right\} ,
\label{sym_basis}
\end{split}
\end{align}
with the $+$ sign for bosonic molecules and the $-$ sign for fermionic molecules.
For the indistinguishable hyperfine kets this relation immediately ensures that only even partial waves are allowed for indistinguishable bosons and only odd partial waves for indistinguishable fermions. In the present case of fermionic NO molecules, it means that a totally anti-symmetric wave function can be expanded onto odd $L$ components.

The total wave function $\Psi(\vec{R})$ is represented as a column vector having the $n$-th component of the form
\begin{eqnarray*}
\Psi_n(\vec{R}) \!=\! \frac{\psi_n(R)}{R}\Big[|{\tilde F}_1,M_{F1};p_1\rangle\!\rangle\!|{\tilde F}_2,M_{F2};p_2\rangle\!\rangle\!|L,M_L\rangle_S\Big]_n,
\end{eqnarray*}
where $\psi_n$ is the diabatic solution of the set of coupled radial equations
\begin{eqnarray}
\left[\sum_{m=1}^{N_{\rm ch}} \left(\!-\frac{\hbar^2}{2m_{\rm red}}\frac{{\rm d}^2}{{\rm d}R^2} + E_m\right)\!\delta_{nm} + V_{nm} \right]\! \psi_m = E_{\rm tot} \psi_n
\label{set_SE}
\end{eqnarray}
with $E_m$ being the threshold energy of the pair of molecules in channel $m$, and $E_{\rm tot} = E_{\rm c} + E_n$, is the total energy written in terms of the collision  energy $E_{\rm c}$ above the threshold energy of channel $n$. Here $V$ includes the dipole-dipole interaction, whose matrix elements are described elsewhere \cite{Avdeenkov05}; and the centrifugal energy, which is diagonal with matrix elements $\hbar^2L(L+1)/(2m_{\rm red}R^2)$.
This Hamiltonian preserves the projection of the total molecular angular momentum on the field axis, $M_{\rm tot}=M_{F1}+M_{F2}+M_L$.

The set of coupled Schr\"odinger equations (\ref{set_SE}) in multichannel scattering is solved using the log-derivative propagator method \cite{Johnson}. Matching the log-derivative matrix with the asymptotic solution for open channels at large $R$ yields the open-open submatrix of the reaction $K$ matrix, and subsequently the scattering matrices $S$ and $T$. For a given incident channel $i$, we identify the partial wave cross sections
\begin{eqnarray*}
\sigma_{L,i \rightarrow f}(E) \!=\! 2 \times \frac{\pi}{k_i^2}\sum_{M_L}\sum_{f,L',M_L'} \big|\langle i,L,M_L |T| f,L',M_L'\rangle\big|^2 \,,
\end{eqnarray*}
where $f=i$ for elastic scattering and $f \ne i$ for inelastic scattering, and the factor of two is required for indistinguishable collision partners.  These cross sections, in turn, are added to yield total cross sections averaged over all incident collision directions,
\begin{eqnarray*}
\sigma_{i \rightarrow f}(E) \!=\! \sum_{L}\sigma_{L,i \rightarrow f}(E)\,.
\end{eqnarray*}

\begin{figure}[hb]
\includegraphics[width=0.47\textwidth]{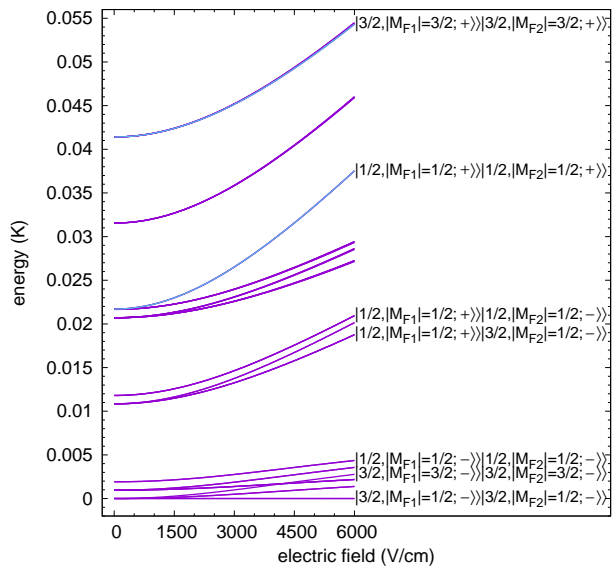}
\caption{Threshold energies for two NO collisional molecules related to the $|3/2,|M_{F1}|=1/2;-\rangle\!\rangle|3/2,|M_{F2}|=1/2;-\rangle\!\rangle_S$ lowest threshold. The light lines indicate the collision channels of our interest, the $|1/2,|M_{F1}|=1/2;+\rangle\!\rangle|1/2,|M_{F2}|=1/2;+\rangle\!\rangle_S$ and $|3/2,|M_{F1}|=3/2;+\rangle\!\rangle|3/2,|M_{F2}|=3/2;+\rangle\!\rangle_S$ channels.}
\label{thr_energies}
\end{figure}


\section{Results and Discussion}

The main goal of the present work is to predict theoretically whether evaporative cooling of $^{14}$N$^{16}$O molecules in their ground rovibrational $^2\Pi_{1/2}$ state trapped in electrostatic traps is feasible. To this end, we consider the weak-field-seeking states, namely, those whose energy rises with electric field in Figure \ref{Starkenergies}.  The molecules in this case have the quantum numbers
$|{\tilde F}M_F;p \rangle$ = $|3/2,1/2;+ \rangle $ and $|3/2,3/2,+ \rangle$.  Scattering calculations shown below are computed for these states using partial waves up to $L=5$, which converge the shown cross sections within approximately 20 percent, adequate for our present purposes.

Based on models similar to the one we use here, cold collisions of $^2\Pi_{3/2}$ molecules have been studied previously \cite{Avdeenkov02,Avdeenkov05,Ticknor05,Quemener13}.  Generally speaking, elastic cross sections can be made large in these states by applying an electric field, given the generically high cross sections of dipolar scattering.  However, it was also found that inelastic scattering grows in an electric field, as the molecules, once polarized, can exert torques on one another, thereby changing their orientation and driving inelastic collisions.

Prospects for mitigating this loss have been considered.  For $^2\Pi_{3/2}$ molecules, it was found that applying a {\it magnetic} field could drive the rates down, partly by decreasing the effective coupling between entrance and exit channels \cite{Ticknor05}, and partly due to a novel long-range shielding potential \cite{Stuhl12}.  This mechanism is, however, ineffective in the $^2\Pi_{1/2}$ state, which has a negligible magnetic moment.  For $\Sigma$ molecules, an additional shielding mechanism is predicted to occur when the energy of the incident threshold crosses the energy of another threshold of states with opposite parity \cite{Avdeenkov06,Wang15_NJP,Quemener16,Martinez17_preprint}. Looking at the thresholds of interest for NO, however (Figure \ref{thr_energies}), this event seems not to occur at realistic electric field values.  

Thus the only hope for suppressing inelastic collisions in the $^2\Pi_{1/2}$ state of NO seems to be to exploit the fact that this molecule is a fermion.  This means that identical NO molecules will necessarily collide with incident partial wave $L=1$, in which case the centrifugal barrier associated with this channel can provide a certain shielding against inelastic scattering.  This shielding mechanism would be effective below the $p$-wave centrifugal barrier, about 4 mK in zero field.  

\begin{figure}[h]
\includegraphics[width=0.47\textwidth]{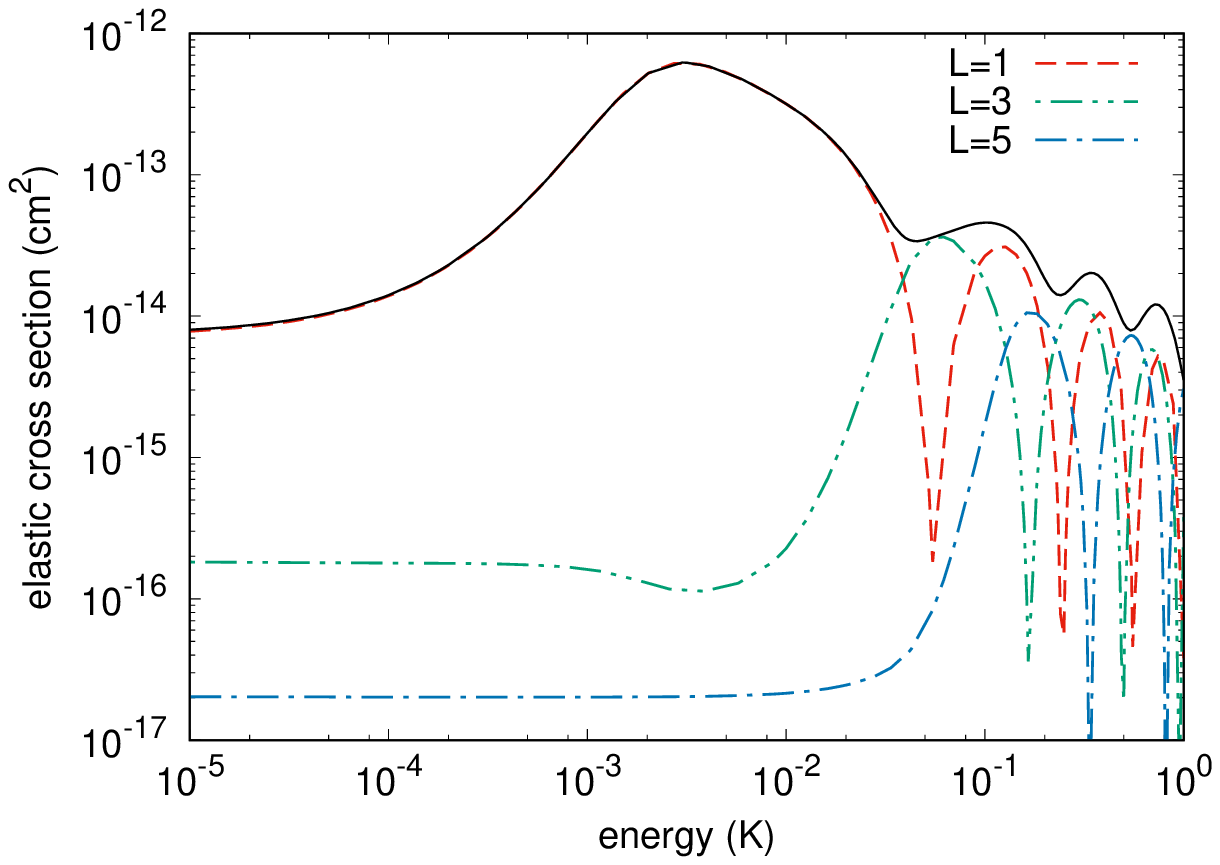}
\includegraphics[width=0.47\textwidth]{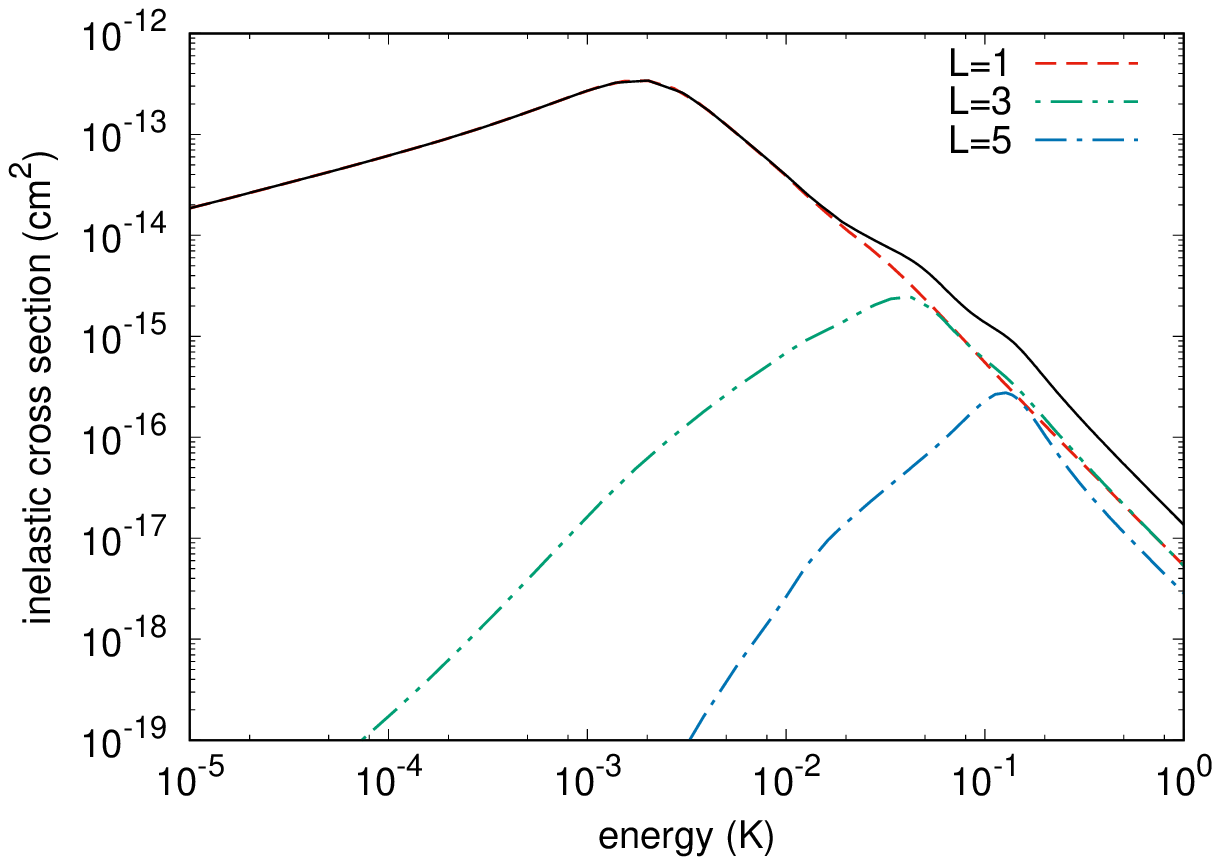}
\caption{ Elastic (upper) and inelastic (lower) cross sections versus the collision energy for 6\,000 V/cm applied electric field , showing individual partial wave contributions $L=1,3,5$ for incident channel $|1/2,1/2;+\rangle\!\rangle|1/2,1/2;+\rangle\!\rangle|1,0\rangle_S$.  The solid line denotes the full cross section.}
\label{cross}
\end{figure}

Examples of total and partial cross sections, as subjected to an electric field of 6\,000 V/cm are shown in Fig. \ref{cross}. Elastic cross sections are shown in the upper panel, while inelastic cross sections are shown in the lower panel.  The collision energy within the wide range 10 $\mu$K through 1 K is considered; note that Stark deceleration down to $\sim 100$ mK temperatures is possible experimentally. At energies below $\sim 1$ mK, cross sections scale according to the usual Wigner threshold laws. Elastic cross sections are independent of energy in all partial waves, as appropriate to dipolar scattering; while inelastic cross sections scale with collision energy $E_c$ as $\sigma_{\rm inel} \sim E_c^{L-1/2}$. Thus elastic scattering will eventually dominate inelastic scattering at sufficiently low temperature, but Figure \ref{cross} shows that $10^{-5}$ K is not yet low enough.

At temperatures above 10 mK, the energy dependence of the cross sections is different.  For elastic scattering of dipoles, once the energy is above the threshold region and several partial waves contribute, the cross section is expected to scale according to the Eikonal approximation, $\sigma_{\rm el} \sim 1/\sqrt{E_c}$ \cite{Bohn09_NJP}. This behavior is seen to approximately hold, even in the energy range shown where a small number of partial waves is relevant and the cross sections exhibit Ramsauer-Townsed minima in each partial wave.  In this energy range the inelastic cross section decreases even more rapidly, as $\sigma_{\rm inel} \sim 1/E_c^2$.  This behavior is different from the $1/E_c^{2/3}$ scaling predicted by the Langevin capture model.  It therefore suggests that the mechanism for state-changing collisions does not require the colliding pair to surmount the centrifugal barriers.  Indeed, we will see below that the inelastic scattering mechanism is due to long-range interactions, captured by the Born approximation.

In any event, elastic scattering is seen to exceed inelastic scattering at collisions energies well above 10 mK.  This temperature therefore presents a bottleneck for evaporative cooling from Stark decelerator temperatures down to the threshold regime.  

\begin{figure}[h]
\includegraphics[width=0.47\textwidth]{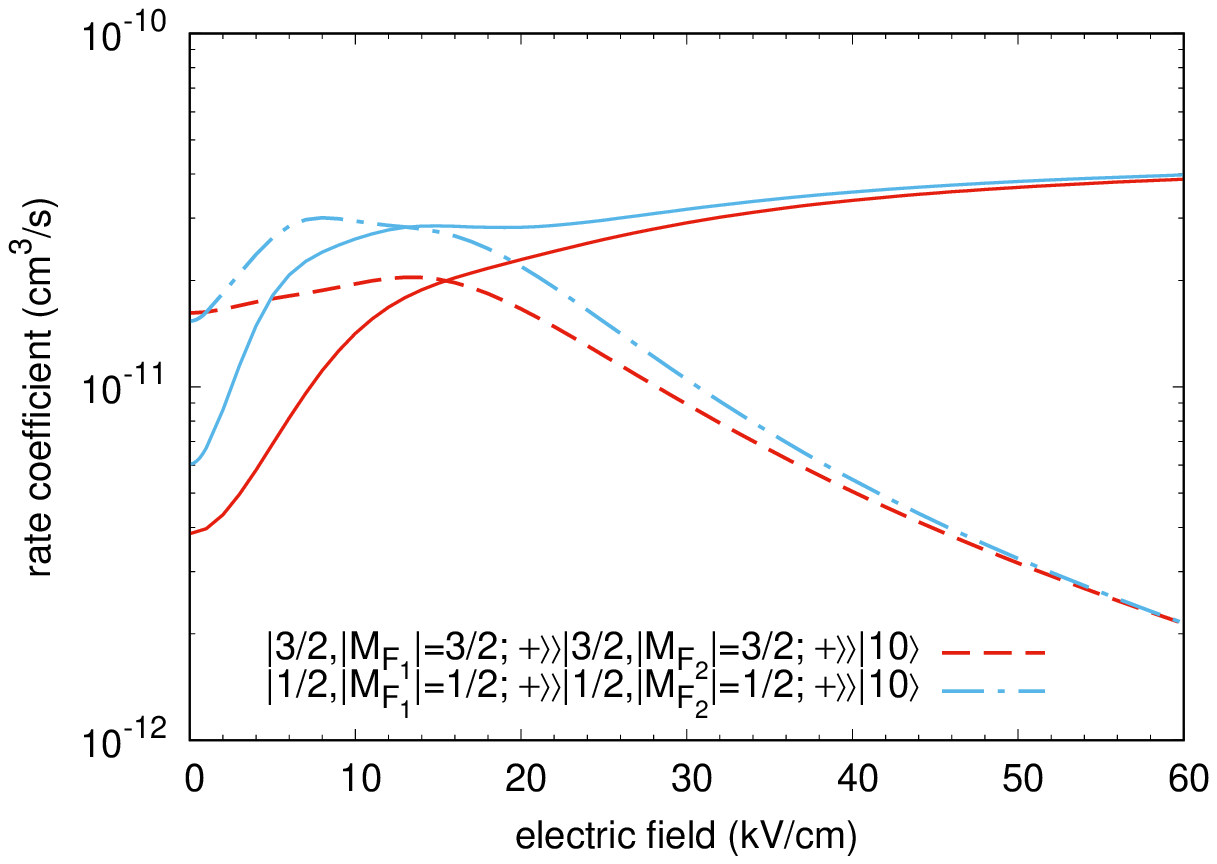}
\includegraphics[width=0.47\textwidth]{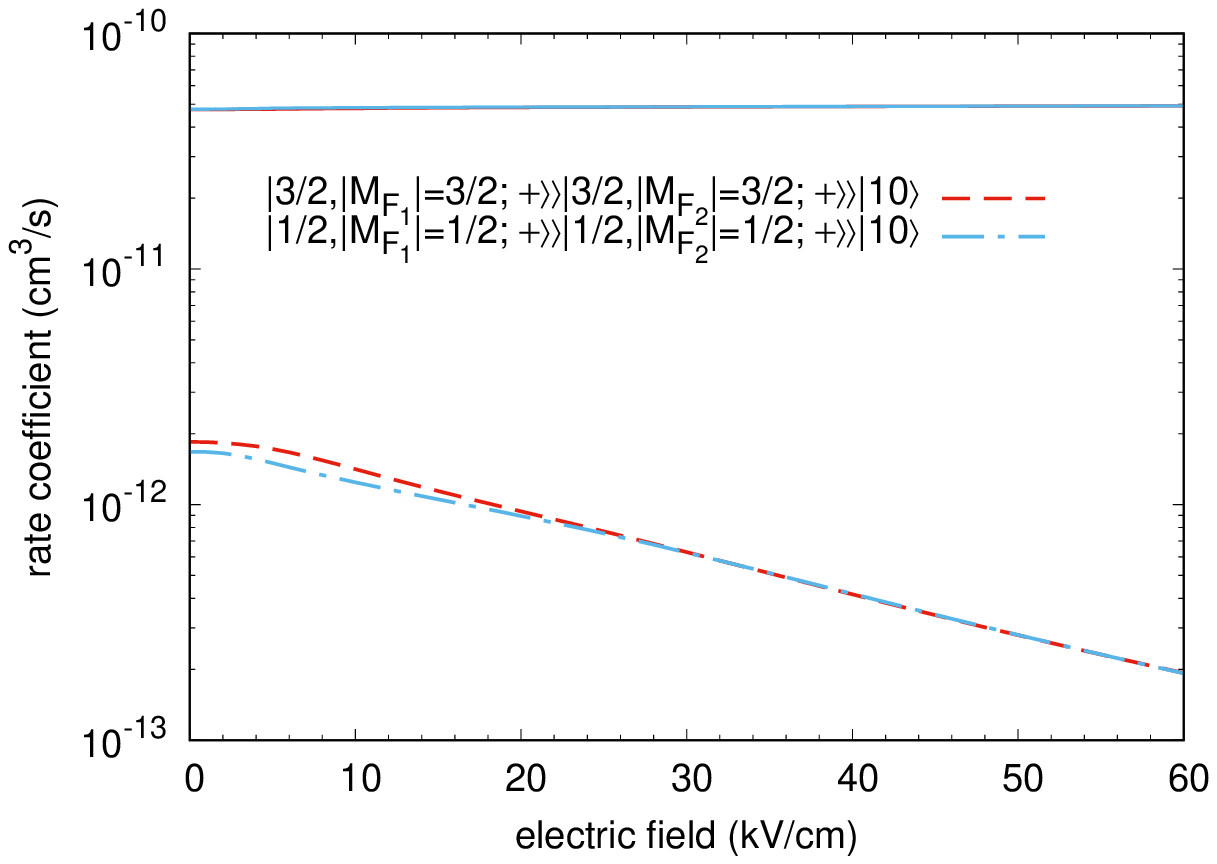}
\caption{Rate coefficients for elastic (solid curves) and inelastic (dashed curves) scattering as a function of electric field for the  two different incident channels considered in the text. The collision energy is fixed at the value $E_{\rm c} = 1$ mK (upper panel) and $E_{\rm c} = 100$ mK. (lower panel)}
\label{rateE10-31}
\end{figure}

It is worthwhile to study whether an electric field can alter the $\sigma_{\rm el}/\sigma_{\rm inel}$ ratio.  Accordingly, Fig. \ref{rateE10-31} shows the elastic and inelastic rate constants, defined by $K = v_i \sigma$ in terms of the incident velocity $v_i$, as a function of electric field.  Both low energy (1 mK,  upper panel) and a higher energy (100 mK, lower panel) are shown.  After an initial rise of inelastic rates, a new effect appears, namely, that the inelastic rate coefficients decrease in large fields. At these higher fields the elastic rates remain high and constant, as expected of the scattering of polarized dipoles.

This dependence on electric field at high field can be seen in the Born approximation.  Consider scattering from an incident channel with wave number $k_i$, to a distinct final channel with wave number $k_f$.  The $T$-matrix element giving the probability amplitude of the transition is \cite{Avdeenkov05}
\begin{eqnarray}
\langle i|T|f\rangle =&& 2 \left( \frac{ 2m_{red} }{ \hbar^2 } \right) \sqrt{k_i k_f} \nonumber \\
&&\times 
\int_0^{\infty} R^2 dR j_{L_i}(k_iR)  
\frac{ C_{i,f} }{ R^3 } j_{L_f}(k_fR),
\label{iTf} \nonumber \\
\end{eqnarray}
where $L_i$, $L_f$ are the partial waves of the incident and final channels; $j_L$'s are spherical Bessel functions; and $C_{if}/R^3$ is the matrix element of the long-range dipolar interaction between the two channels.  Note that $C_{if}$ is, in general, a function of electric field, since the channel indices refer to molecular states as dressed by the electric field.  
Working out the integral, the $T$-matrix elements are 
\begin{eqnarray}
\langle i | T | f \rangle =&&
\frac{ 2 \pi m_{\rm red} C_{if} }{ \hbar^2 }
\left[ \frac{ \Gamma \left( \frac{ L_i + L_f }{ 2 } \right) }
{ 4 \Gamma \left( \frac{ -L_i + L_f + 3 }{ 2 } \right) \Gamma(L_i+3/2) } \right]
\nonumber \\
&& \times \frac{ k_i^{L_i + 1/2} }{ k_f^{L_f-1/2} } \\
&& \times 
F \left( \frac{ L_i + L_f }{ 2 }, \frac{ L_i - L_f - 1 }{ 2 }; L_i+3/2;
\left( \frac{ k_i }{ k_f } \right)^2 \right), \nonumber
\end{eqnarray}
where $F$ is a hypergeometric function.

For ultracold collisions, $k_i$ remains small, while $k_f$ continues to grow with electric field.  Quantitatively, the outgoing wave number scales with the field ${\cal E}$, for large field, as 
\begin{eqnarray}
k_f &=& \sqrt{ (2m_{\rm red}/\hbar^2) ( E_{\rm tot} - E_f )} \nonumber \\
&\propto& \sqrt{ d {\cal E} },
\end{eqnarray}
given that the threshold energy is proportional to the field in the linear Stark regime.
Moreover, in this limit the argument $(k_i/k_f)^2$ of the hypergeometric function vanishes, whereby $F=1$ and for $L_f=1$ final channels, the ultracold inelastic rate should scale as $K_{\rm inel} \propto 1/k_f \propto 1/\sqrt{ {\cal E }}$. 

\begin{figure}[h]
\includegraphics[width=0.47\textwidth]{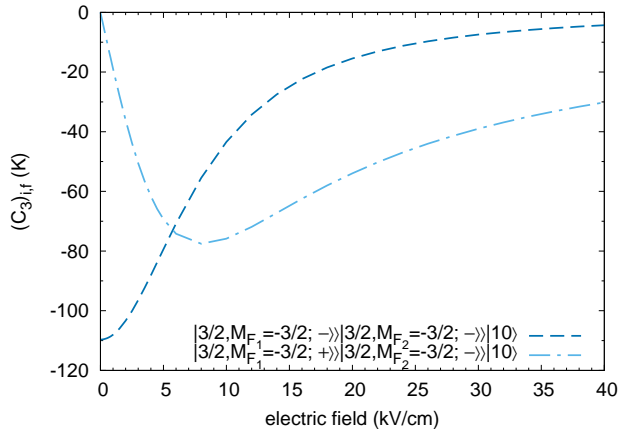}
\caption{Off-diagonal matrix elements $C_{i,f}$ of dipole-dipole interaction without a radial dependence as functions of electric field for outbound channels indicated in the legend and inbound channel $|3/2,-3/2;+\rangle\!\rangle|3/2,-3/2;+\rangle\!\rangle|1,0\rangle_S$.}
\label{C3}
\end{figure}

This behavior is approached in the field regime where the molecules are fully polarized and the potential couplings $C_{if}$ saturate with field.  As shown in Figure \ref{C3}, this saturation has not yet quite occurred for fields as large as 40 kV/cm.  In any event, the suppression effect would require fields greater than 10\,kV/cm, which are likely unfeasible in electrostatic traps.

Finally, we note that we have carried out scattering calculations under the influence of a magnetic field.  The cross sections show negligible dependence on magnetic field up to 1000 G, owing to the small $g$-factor of the $^2\Pi_{1/2}$ state.  A greater influence of magnetic field may be expected for the metastable $^2\Pi_{3/2}, J=3/2$ state.  This remains a subject of future investigations.


\section{Conclusion}
We have performed analysis of possibility of evaporative cooling of NO molecules in their ground electronic $\Omega=1/2$ state and computed scattering cross sections and rate coefficients under influence of electric fields. 
We find that the ratio of elastic to inelastic rates is favorable at higher temperatures $\sim 0.5$\,K and it appears that applying an electric field can improve the prospects for evaporative cooling even at 100\,mK. However, the required fields are probably unrealistically high with the traps as currently built.

\section{Acknowledgments}

We acknowledge funding from the U.S. Army Research Office under ARO Grant No. W911NF-12-1-0476, and useful discussions with B. van der Meerakker.

\bibliography{Bibliography}

\begin{thebibliography}{32}
\expandafter\ifx\csname natexlab\endcsname\relax\def\natexlab#1{#1}\fi
\expandafter\ifx\csname bibnamefont\endcsname\relax
  \def\bibnamefont#1{#1}\fi
\expandafter\ifx\csname bibfnamefont\endcsname\relax
  \def\bibfnamefont#1{#1}\fi
\expandafter\ifx\csname citenamefont\endcsname\relax
  \def\citenamefont#1{#1}\fi
\expandafter\ifx\csname url\endcsname\relax
  \def\url#1{\texttt{#1}}\fi
\expandafter\ifx\csname urlprefix\endcsname\relax\def\urlprefix{URL }\fi
\providecommand{\bibinfo}[2]{#2}
\providecommand{\eprint}[2][]{\url{#2}}

\bibitem[{\citenamefont{Shuman et~al.}(2010)\citenamefont{Shuman, Barry, and
  DeMille}}]{Shuman10_Nat}
\bibinfo{author}{\bibfnamefont{E.~S.} \bibnamefont{Shuman}},
  \bibinfo{author}{\bibfnamefont{J.~F.} \bibnamefont{Barry}}, \bibnamefont{and}
  \bibinfo{author}{\bibfnamefont{D.}~\bibnamefont{DeMille}},
  \bibinfo{journal}{Nature} \textbf{\bibinfo{volume}{467}},
  \bibinfo{pages}{820} (\bibinfo{year}{2010}).

\bibitem[{\citenamefont{Barry et~al.}(2014)\citenamefont{Barry, McCarron,
  Norrgard, Steinecker, and DeMille}}]{Barry14_Nat}
\bibinfo{author}{\bibfnamefont{J.~F.} \bibnamefont{Barry}},
  \bibinfo{author}{\bibfnamefont{D.~J.} \bibnamefont{McCarron}},
  \bibinfo{author}{\bibfnamefont{E.~B.} \bibnamefont{Norrgard}},
  \bibinfo{author}{\bibfnamefont{M.~H.} \bibnamefont{Steinecker}},
  \bibnamefont{and} \bibinfo{author}{\bibfnamefont{D.}~\bibnamefont{DeMille}},
  \bibinfo{journal}{Nature} \textbf{\bibinfo{volume}{512}},
  \bibinfo{pages}{286} (\bibinfo{year}{2014}).

\bibitem[{\citenamefont{Yeo et~al.}(2015)\citenamefont{Yeo, Hummon, Collopy,
  Yan, Hemmerling, Chae, Doyle, and Ye}}]{Yeo15_PRL}
\bibinfo{author}{\bibfnamefont{M.}~\bibnamefont{Yeo}},
  \bibinfo{author}{\bibfnamefont{M.~T.} \bibnamefont{Hummon}},
  \bibinfo{author}{\bibfnamefont{A.~L.} \bibnamefont{Collopy}},
  \bibinfo{author}{\bibfnamefont{B.}~\bibnamefont{Yan}},
  \bibinfo{author}{\bibfnamefont{B.}~\bibnamefont{Hemmerling}},
  \bibinfo{author}{\bibfnamefont{E.}~\bibnamefont{Chae}},
  \bibinfo{author}{\bibfnamefont{J.~M.} \bibnamefont{Doyle}}, \bibnamefont{and}
  \bibinfo{author}{\bibfnamefont{J.}~\bibnamefont{Ye}}, \bibinfo{journal}{Phys.
  Rev. Lett.} \textbf{\bibinfo{volume}{114}}, \bibinfo{pages}{223003}
  (\bibinfo{year}{2015}).

\bibitem[{\citenamefont{Kozyryev et~al.}(2016)\citenamefont{Kozyryev, Baum,
  Matsuda, and Doyle}}]{Kozyryev16_CPC}
\bibinfo{author}{\bibfnamefont{I.}~\bibnamefont{Kozyryev}},
  \bibinfo{author}{\bibfnamefont{L.}~\bibnamefont{Baum}},
  \bibinfo{author}{\bibfnamefont{K.}~\bibnamefont{Matsuda}}, \bibnamefont{and}
  \bibinfo{author}{\bibfnamefont{J.~M.} \bibnamefont{Doyle}},
  \bibinfo{journal}{Chem. Phys. Chem} \textbf{\bibinfo{volume}{17}},
  \bibinfo{pages}{3641} (\bibinfo{year}{2016}).

\bibitem[{\citenamefont{Chae et~al.}(2017)\citenamefont{Chae, Anderegg,
  Augenbraun, Ravi, Hemmerling, Hutzler, Collopy, Ye, Ketterle, and
  Doyle}}]{Chae17_NJP}
\bibinfo{author}{\bibfnamefont{E.}~\bibnamefont{Chae}},
  \bibinfo{author}{\bibfnamefont{L.}~\bibnamefont{Anderegg}},
  \bibinfo{author}{\bibfnamefont{B.~L.} \bibnamefont{Augenbraun}},
  \bibinfo{author}{\bibfnamefont{A.}~\bibnamefont{Ravi}},
  \bibinfo{author}{\bibfnamefont{B.}~\bibnamefont{Hemmerling}},
  \bibinfo{author}{\bibfnamefont{N.~R.} \bibnamefont{Hutzler}},
  \bibinfo{author}{\bibfnamefont{A.~L.} \bibnamefont{Collopy}},
  \bibinfo{author}{\bibfnamefont{J.}~\bibnamefont{Ye}},
  \bibinfo{author}{\bibfnamefont{W.}~\bibnamefont{Ketterle}}, \bibnamefont{and}
  \bibinfo{author}{\bibfnamefont{J.~M.} \bibnamefont{Doyle}},
  \bibinfo{journal}{New Journal of Physics} \textbf{\bibinfo{volume}{19}},
  \bibinfo{pages}{033035} (\bibinfo{year}{2017}).

\bibitem[{\citenamefont{Anderegg et~al.}(2017)\citenamefont{Anderegg,
  Augenbraun, Chae, Hemmerling, Hutzler, Ravi, Collopy, Ye, Ketterle, and
  Doyle}}]{Anderegg17_preprint}
\bibinfo{author}{\bibfnamefont{L.}~\bibnamefont{Anderegg}},
  \bibinfo{author}{\bibfnamefont{B.}~\bibnamefont{Augenbraun}},
  \bibinfo{author}{\bibfnamefont{E.}~\bibnamefont{Chae}},
  \bibinfo{author}{\bibfnamefont{B.}~\bibnamefont{Hemmerling}},
  \bibinfo{author}{\bibfnamefont{N.~R.} \bibnamefont{Hutzler}},
  \bibinfo{author}{\bibfnamefont{A.}~\bibnamefont{Ravi}},
  \bibinfo{author}{\bibfnamefont{A.}~\bibnamefont{Collopy}},
  \bibinfo{author}{\bibfnamefont{J.}~\bibnamefont{Ye}},
  \bibinfo{author}{\bibfnamefont{W.}~\bibnamefont{Ketterle}}, \bibnamefont{and}
  \bibinfo{author}{\bibfnamefont{J.}~\bibnamefont{Doyle}}
  (\bibinfo{year}{2017}), \bibinfo{note}{arXiv:1705.10288}.

\bibitem[{\citenamefont{Hutzler et~al.}(2012)\citenamefont{Hutzler, Lu, and
  Doyle}}]{Hutzler12_ChemRev}
\bibinfo{author}{\bibfnamefont{N.~R.} \bibnamefont{Hutzler}},
  \bibinfo{author}{\bibfnamefont{H.-I.} \bibnamefont{Lu}}, \bibnamefont{and}
  \bibinfo{author}{\bibfnamefont{J.~M.} \bibnamefont{Doyle}},
  \bibinfo{journal}{Chemical Reviews} \textbf{\bibinfo{volume}{112}},
  \bibinfo{pages}{4803} (\bibinfo{year}{2012}).

\bibitem[{\citenamefont{Ospelkaus et~al.}(2010)\citenamefont{Ospelkaus, Ni,
  Wang, de~Miranda, Neyenhuis, Qu{\'e}m{\'e}ner, Julienne, Bohn, Jin, and
  Ye}}]{Ospelkaus10_Sci}
\bibinfo{author}{\bibfnamefont{S.}~\bibnamefont{Ospelkaus}},
  \bibinfo{author}{\bibfnamefont{K.-K.} \bibnamefont{Ni}},
  \bibinfo{author}{\bibfnamefont{D.}~\bibnamefont{Wang}},
  \bibinfo{author}{\bibfnamefont{M.~H.~G.} \bibnamefont{de~Miranda}},
  \bibinfo{author}{\bibfnamefont{B.}~\bibnamefont{Neyenhuis}},
  \bibinfo{author}{\bibfnamefont{G.}~\bibnamefont{Qu{\'e}m{\'e}ner}},
  \bibinfo{author}{\bibfnamefont{P.~S.} \bibnamefont{Julienne}},
  \bibinfo{author}{\bibfnamefont{J.~L.} \bibnamefont{Bohn}},
  \bibinfo{author}{\bibfnamefont{D.~S.} \bibnamefont{Jin}}, \bibnamefont{and}
  \bibinfo{author}{\bibfnamefont{J.}~\bibnamefont{Ye}},
  \bibinfo{journal}{Science} \textbf{\bibinfo{volume}{327}},
  \bibinfo{pages}{853} (\bibinfo{year}{2010}).

\bibitem[{\citenamefont{Monroe et~al.}(1993)\citenamefont{Monroe, Cornell,
  Sackett, Myatt, and Wieman}}]{Monroe93_PRL}
\bibinfo{author}{\bibfnamefont{C.~R.} \bibnamefont{Monroe}},
  \bibinfo{author}{\bibfnamefont{E.~A.} \bibnamefont{Cornell}},
  \bibinfo{author}{\bibfnamefont{C.~A.} \bibnamefont{Sackett}},
  \bibinfo{author}{\bibfnamefont{C.~J.} \bibnamefont{Myatt}}, \bibnamefont{and}
  \bibinfo{author}{\bibfnamefont{C.~E.} \bibnamefont{Wieman}},
  \bibinfo{journal}{Phys. Rev. Lett.} \textbf{\bibinfo{volume}{70}},
  \bibinfo{pages}{414} (\bibinfo{year}{1993}).

\bibitem[{\citenamefont{Stuhl et~al.}(2012)\citenamefont{Stuhl, Hummon, Yeo,
  Qu\'em\'ener, Bohn, and Ye}}]{Stuhl12}
\bibinfo{author}{\bibfnamefont{B.~K.} \bibnamefont{Stuhl}},
  \bibinfo{author}{\bibfnamefont{M.~T.} \bibnamefont{Hummon}},
  \bibinfo{author}{\bibfnamefont{M.}~\bibnamefont{Yeo}},
  \bibinfo{author}{\bibfnamefont{G.}~\bibnamefont{Qu\'em\'ener}},
  \bibinfo{author}{\bibfnamefont{J.~L.} \bibnamefont{Bohn}}, \bibnamefont{and}
  \bibinfo{author}{\bibfnamefont{J.}~\bibnamefont{Ye}},
  \bibinfo{journal}{Nature} \textbf{\bibinfo{volume}{492}},
  \bibinfo{pages}{396} (\bibinfo{year}{2012}).

\bibitem[{\citenamefont{Reens et~al.}(2017)\citenamefont{Reens, Wu, Langen, and
  Ye}}]{Reens17_preprint}
\bibinfo{author}{\bibfnamefont{D.}~\bibnamefont{Reens}},
  \bibinfo{author}{\bibfnamefont{H.}~\bibnamefont{Wu}},
  \bibinfo{author}{\bibfnamefont{T.}~\bibnamefont{Langen}}, \bibnamefont{and}
  \bibinfo{author}{\bibfnamefont{J.}~\bibnamefont{Ye}} (\bibinfo{year}{2017}),
  \bibinfo{note}{arXiv:1706.02806}.

\bibitem[{\citenamefont{Balakrishnan}(2016)}]{Bala16_JCP}
\bibinfo{author}{\bibfnamefont{N.}~\bibnamefont{Balakrishnan}},
  \bibinfo{journal}{The Journal of Chemical Physics}
  \textbf{\bibinfo{volume}{145}}, \bibinfo{pages}{150901}
  (\bibinfo{year}{2016}).

\bibitem[{\citenamefont{Vogels et~al.}(2015)\citenamefont{Vogels, Onvlee,
  Chefdeville, van~der Avoird, Groenenboom, and van~de
  Meerakker}}]{Vogels15_Sci}
\bibinfo{author}{\bibfnamefont{S.~N.} \bibnamefont{Vogels}},
  \bibinfo{author}{\bibfnamefont{J.}~\bibnamefont{Onvlee}},
  \bibinfo{author}{\bibfnamefont{S.}~\bibnamefont{Chefdeville}},
  \bibinfo{author}{\bibfnamefont{A.}~\bibnamefont{van~der Avoird}},
  \bibinfo{author}{\bibfnamefont{G.~C.} \bibnamefont{Groenenboom}},
  \bibnamefont{and} \bibinfo{author}{\bibfnamefont{S.~Y.~T.}
  \bibnamefont{van~de Meerakker}}, \bibinfo{journal}{Science}
  \textbf{\bibinfo{volume}{350}}, \bibinfo{pages}{787} (\bibinfo{year}{2015}).

\bibitem[{\citenamefont{Bichsel et~al.}(2007)\citenamefont{Bichsel, Morrison,
  Shafer-Ray, and Abraham}}]{Bichsel07_PRA}
\bibinfo{author}{\bibfnamefont{B.~J.} \bibnamefont{Bichsel}},
  \bibinfo{author}{\bibfnamefont{M.~A.} \bibnamefont{Morrison}},
  \bibinfo{author}{\bibfnamefont{N.}~\bibnamefont{Shafer-Ray}},
  \bibnamefont{and} \bibinfo{author}{\bibfnamefont{E.~R.~I.}
  \bibnamefont{Abraham}}, \bibinfo{journal}{Phys. Rev. A}
  \textbf{\bibinfo{volume}{75}}, \bibinfo{pages}{023410}
  (\bibinfo{year}{2007}).

\bibitem[{\citenamefont{Elioff et~al.}(2003)\citenamefont{Elioff, Valentini,
  and Chandler}}]{Elioff03_Sci}
\bibinfo{author}{\bibfnamefont{M.~S.} \bibnamefont{Elioff}},
  \bibinfo{author}{\bibfnamefont{J.~J.} \bibnamefont{Valentini}},
  \bibnamefont{and} \bibinfo{author}{\bibfnamefont{D.~W.}
  \bibnamefont{Chandler}}, \textbf{\bibinfo{volume}{302}},
  \bibinfo{pages}{1940} (\bibinfo{year}{2003}).

\bibitem[{\citenamefont{Varberg et~al.}(1999)\citenamefont{Varberg, Stroh, and
  Evenson}}]{Varberg}
\bibinfo{author}{\bibfnamefont{T.~D.} \bibnamefont{Varberg}},
  \bibinfo{author}{\bibfnamefont{F.}~\bibnamefont{Stroh}}, \bibnamefont{and}
  \bibinfo{author}{\bibfnamefont{K.~M.} \bibnamefont{Evenson}},
  \bibinfo{journal}{Journal of Molecular Spectroscopy}
  \textbf{\bibinfo{volume}{196}}, \bibinfo{pages}{5 } (\bibinfo{year}{1999}).

\bibitem[{\citenamefont{Meerts and Dymanus}(1972)}]{Meerts}
\bibinfo{author}{\bibfnamefont{W.}~\bibnamefont{Meerts}} \bibnamefont{and}
  \bibinfo{author}{\bibfnamefont{A.}~\bibnamefont{Dymanus}},
  \bibinfo{journal}{Journal of Molecular Spectroscopy}
  \textbf{\bibinfo{volume}{44}}, \bibinfo{pages}{320 } (\bibinfo{year}{1972}).

\bibitem[{\citenamefont{Avdeenkov and Bohn}(2002)}]{Avdeenkov02}
\bibinfo{author}{\bibfnamefont{A.~V.} \bibnamefont{Avdeenkov}}
  \bibnamefont{and} \bibinfo{author}{\bibfnamefont{J.~L.} \bibnamefont{Bohn}},
  \bibinfo{journal}{Phys. Rev. A} \textbf{\bibinfo{volume}{66}},
  \bibinfo{pages}{052718} (\bibinfo{year}{2002}).

\bibitem[{\citenamefont{Ticknor and Bohn}(2005)}]{Ticknor05}
\bibinfo{author}{\bibfnamefont{C.}~\bibnamefont{Ticknor}} \bibnamefont{and}
  \bibinfo{author}{\bibfnamefont{J.~L.} \bibnamefont{Bohn}},
  \bibinfo{journal}{Phys. Rev. A} \textbf{\bibinfo{volume}{71}},
  \bibinfo{pages}{022709} (\bibinfo{year}{2005}).

\bibitem[{\citenamefont{Lide}(2003)}]{CRC}
\bibinfo{author}{\bibfnamefont{D.~R.} \bibnamefont{Lide}},
  \emph{\bibinfo{title}{CRC Handbook of Chemisty and Physics, 84th Edition}}
  (\bibinfo{publisher}{CRC Press}, \bibinfo{year}{2003}).

\bibitem[{\citenamefont{Mizushima et~al.}(1955)\citenamefont{Mizushima, Cox,
  and Gordy}}]{Mizushima55_PR}
\bibinfo{author}{\bibfnamefont{M.}~\bibnamefont{Mizushima}},
  \bibinfo{author}{\bibfnamefont{J.~T.} \bibnamefont{Cox}}, \bibnamefont{and}
  \bibinfo{author}{\bibfnamefont{W.}~\bibnamefont{Gordy}},
  \bibinfo{journal}{Phys. Rev.} \textbf{\bibinfo{volume}{98}},
  \bibinfo{pages}{1034} (\bibinfo{year}{1955}).

\bibitem[{\citenamefont{Zhou}(1993)}]{Zhou}
\bibinfo{author}{\bibfnamefont{G.-D.} \bibnamefont{Zhou}},
  \emph{\bibinfo{title}{Fundamentals of Structural Chemistry}}
  (\bibinfo{publisher}{World Scientific Publishing Co Pte Ltd},
  \bibinfo{address}{Singapore}, \bibinfo{year}{1993}).

\bibitem[{\citenamefont{Bodo et~al.}(2004)\citenamefont{Bodo, Gianturco,
  Balakrishnan, and Dalgarno}}]{Bodo04_JPB}
\bibinfo{author}{\bibfnamefont{E.}~\bibnamefont{Bodo}},
  \bibinfo{author}{\bibfnamefont{F.~A.} \bibnamefont{Gianturco}},
  \bibinfo{author}{\bibfnamefont{N.}~\bibnamefont{Balakrishnan}},
  \bibnamefont{and} \bibinfo{author}{\bibfnamefont{A.}~\bibnamefont{Dalgarno}},
  \bibinfo{journal}{Journal of Physics B: Atomic, Molecular and Optical
  Physics} \textbf{\bibinfo{volume}{37}}, \bibinfo{pages}{3641}
  (\bibinfo{year}{2004}).

\bibitem[{\citenamefont{Bridge and Buckingham}(1966)}]{Bridge}
\bibinfo{author}{\bibfnamefont{N.~J.} \bibnamefont{Bridge}} \bibnamefont{and}
  \bibinfo{author}{\bibfnamefont{A.~D.} \bibnamefont{Buckingham}},
  \bibinfo{journal}{Proceedings of the Royal Society of London A: Mathematical,
  Physical and Engineering Sciences} \textbf{\bibinfo{volume}{295}},
  \bibinfo{pages}{334} (\bibinfo{year}{1966}).

\bibitem[{\citenamefont{Avdeenkov and Bohn}(2005)}]{Avdeenkov05}
\bibinfo{author}{\bibfnamefont{A.~V.} \bibnamefont{Avdeenkov}}
  \bibnamefont{and} \bibinfo{author}{\bibfnamefont{J.~L.} \bibnamefont{Bohn}},
  \bibinfo{journal}{Phys. Rev. A} \textbf{\bibinfo{volume}{71}},
  \bibinfo{pages}{022706} (\bibinfo{year}{2005}).

\bibitem[{\citenamefont{Johnson}(1973)}]{Johnson}
\bibinfo{author}{\bibfnamefont{B.~R.} \bibnamefont{Johnson}},
  \bibinfo{journal}{Journal of Computational Physics}
  \textbf{\bibinfo{volume}{13}}, \bibinfo{pages}{445 } (\bibinfo{year}{1973}).

\bibitem[{\citenamefont{Qu\'em\'ener and Bohn}(2013)}]{Quemener13}
\bibinfo{author}{\bibfnamefont{G.}~\bibnamefont{Qu\'em\'ener}}
  \bibnamefont{and} \bibinfo{author}{\bibfnamefont{J.~L.} \bibnamefont{Bohn}},
  \bibinfo{journal}{Phys. Rev. A} \textbf{\bibinfo{volume}{88}},
  \bibinfo{pages}{012706} (\bibinfo{year}{2013}).

\bibitem[{\citenamefont{Avdeenkov et~al.}(2006)\citenamefont{Avdeenkov, Kajita,
  and Bohn}}]{Avdeenkov06}
\bibinfo{author}{\bibfnamefont{A.~V.} \bibnamefont{Avdeenkov}},
  \bibinfo{author}{\bibfnamefont{M.}~\bibnamefont{Kajita}}, \bibnamefont{and}
  \bibinfo{author}{\bibfnamefont{J.~L.} \bibnamefont{Bohn}},
  \bibinfo{journal}{Phys. Rev. A} \textbf{\bibinfo{volume}{73}},
  \bibinfo{pages}{022707} (\bibinfo{year}{2006}).

\bibitem[{\citenamefont{Wang and Qu\'em\'ener}(2015)}]{Wang15_NJP}
\bibinfo{author}{\bibfnamefont{G.}~\bibnamefont{Wang}} \bibnamefont{and}
  \bibinfo{author}{\bibfnamefont{G.}~\bibnamefont{Qu\'em\'ener}},
  \bibinfo{journal}{New Journal of Physics} \textbf{\bibinfo{volume}{17}},
  \bibinfo{pages}{035015} (\bibinfo{year}{2015}).

\bibitem[{\citenamefont{Qu\'em\'ener and Bohn}(2016)}]{Quemener16}
\bibinfo{author}{\bibfnamefont{G.}~\bibnamefont{Qu\'em\'ener}}
  \bibnamefont{and} \bibinfo{author}{\bibfnamefont{J.~L.} \bibnamefont{Bohn}},
  \bibinfo{journal}{Phys. Rev. A} \textbf{\bibinfo{volume}{93}},
  \bibinfo{pages}{012704} (\bibinfo{year}{2016}).

\bibitem[{\citenamefont{Gonz\'{a}lez-Mart\'{i}nez
  et~al.}(2017)\citenamefont{Gonz\'{a}lez-Mart\'{i}nez, Bohn, and
  Qu\'{e}m\'{e}ner}}]{Martinez17_preprint}
\bibinfo{author}{\bibfnamefont{M.~L.} \bibnamefont{Gonz\'{a}lez-Mart\'{i}nez}},
  \bibinfo{author}{\bibfnamefont{J.~L.} \bibnamefont{Bohn}}, \bibnamefont{and}
  \bibinfo{author}{\bibfnamefont{G.}~\bibnamefont{Qu\'{e}m\'{e}ner}}
  (\bibinfo{year}{2017}), \bibinfo{note}{arXiv:1707.03239}.

\bibitem[{\citenamefont{Bohn et~al.}(2009)\citenamefont{Bohn, Cavagnero, and
  Ticknor}}]{Bohn09_NJP}
\bibinfo{author}{\bibfnamefont{J.~L.} \bibnamefont{Bohn}},
  \bibinfo{author}{\bibfnamefont{M.}~\bibnamefont{Cavagnero}},
  \bibnamefont{and} \bibinfo{author}{\bibfnamefont{C.}~\bibnamefont{Ticknor}},
  \bibinfo{journal}{New Journal of Physics} \textbf{\bibinfo{volume}{11}},
  \bibinfo{pages}{055039} (\bibinfo{year}{2009}).

\end{thebibliography}

\end{document}